\documentclass[aps,prd,floats,floatfix,nofootinbib,twocolumn]{revtex4-1}
\usepackage{graphicx}
\usepackage{amssymb}
\usepackage{hyperref}
\usepackage{xcolor}

\usepackage{amsmath}

\begin{document}

\title{Implications of the observation of dark matter self-interactions\\ for singlet scalar dark matter}

\author{Robyn Campbell}
\email{RobynCampbell@cmail.carleton.ca}

\author{Stephen Godfrey}
\email{godfrey@physics.carleton.ca}

\author{Heather E.~Logan}
\email{logan@physics.carleton.ca} 

\author{Andrea D.~Peterson}
\email{apeterso@physics.carleton.ca}

\author{Alexandre Poulin}
\email{AlexandrePoulin@cmail.carleton.ca}

\affiliation{Ottawa-Carleton Institute for Physics, Carleton University, 1125 Colonel By Drive, Ottawa, Ontario K1S 5B6, Canada}

\date{February 3, 2020}                                  % Activate to display a given date or no date

\begin{abstract}
Evidence for dark matter self-interactions has recently been reported based on the observation of a spatial offset between the dark matter halo and the stars in a galaxy in the cluster Abell 3827.  Interpreting the offset as due to dark matter self-interactions leads to a cross section measurement of $\sigma_{\rm DM}/m \sim (1-1.5)~{\rm cm}^{2}~{\rm g}^{-1} $, where $m$ is the mass of the dark matter particle.  We use this observation to constrain singlet scalar dark matter coupled to the Standard Model and to two-Higgs-doublet models.
We show that the most natural scenario in this class of models is very light dark matter, below about 0.1~GeV, whose relic abundance is set by \emph{freeze-in}, i.e., by slow production of dark matter in the early universe via extremely tiny interactions with the Higgs boson, never reaching thermal equilibrium.  We also show that the dark matter abundance can be established through the usual thermal freeze-out mechanism in the singlet scalar extension of the Yukawa-aligned two-Higgs-doublet model, but that it requires rather severe fine tuning of the singlet scalar mass.
\end{abstract}

\maketitle 

%%%%%%%%%%%%%%%%%%%%%%%%%%%%%%%%%%%%%%%%%%%%%%
\section{Introduction}

Understanding the nature of dark matter is one of the holy grails of modern particle physics.  Up to now, dark matter has been observed only through its gravitational interactions with Standard Model (SM) particles (for a recent review, see, e.g., Ref.~\cite{Gelmini:2015zpa}).

Very recently, the first evidence for dark matter self-interactions has been reported~\cite{Massey:2015oza} based on observations of four elliptical galaxies in the inner 10~kpc core of galaxy cluster Abell 3827.  Using gravitational lensing, Ref.~\cite{Massey:2015oza} reconstructs the dark matter halos of the four galaxies and observes that at least one of the halos is spatially offset from its stars by a distance of $\Delta = 1.62^{+0.47}_{-0.49}~{\rm kpc}$.  Dark matter self-interactions can lead to such an offset: the dark matter halos of the infalling galaxies will be impeded as they pass through the dark matter associated with the cluster. The magnitude of this effect scales with the self-interaction cross section, which can therefore be extracted from the size of the offset. 

Kahlhoefer {\it et al.}~\cite{Kahlhoefer:2015vua} consider the case of dark matter contact interactions, which will be relevant for the scalar dark matter models we consider below. 
Whereas the effect on the halo of frequent dark matter interactions can be described as a drag force, in the case of rare, high-momentum-transfer contact interactions, this picture is inaccurate. Rather, in the contact-interaction scenario the observed separation is due to the evaporation of dark matter from the halo. Most dark matter particles in the halo will not interact, but those that do scatter will be ejected from the halo in the backward direction. Eventually the scattered particles will become substantially separated from the halo, but at short time scales, they remain observationally associated with the galaxy. This leads to a deformation of the halo and an offset between the centroids of the galaxy and the halo. Determination of the size of the offset is fairly sensitive to how the centroid is defined, but Ref.~\cite{Kahlhoefer:2015vua} finds that a cross section of 
\begin{equation}
	\sigma_{\rm DM}/m \sim 1.5~{\rm cm}^{2}~{\rm g}^{-1}
	\label{eq:xsec}
\end{equation} 
could account for the offset observed in Ref.~\cite{Massey:2015oza}. This is in slight tension with results from previous studies of other astrophysical systems~\cite{Kahlhoefer:2015vua,Markevitch:2003at,Peter:2012jh,Randall:2007ph,Rocha:2012jg,Harvey:2015hha}, 
which find an upper bound on the cross section of
\begin{equation}
	\sigma_{\rm DM}/m \lesssim 1~{\rm cm}^{2}~{\rm g}^{-1}. 
	\label{eq:upperbound}
\end{equation}
We will use Eqs.~(\ref{eq:xsec}) and (\ref{eq:upperbound}) to set a range for the estimated self-interaction cross section. In particle physics units, this corresponds to
\begin{equation}
	\sigma_{\rm DM}/m \sim (4.7-7.0) \times 10^3~{\rm GeV}^{-3}.
	\label{eq:xsec2}
\end{equation}  

 Of course, it is possible that the offset may be an astrophysical artifact. Nevertheless, if a dark matter self-interaction interpretation holds up, it would represent a profound advance in our knowledge of the nature of dark matter.  It is therefore of great importance to explore its implications for models of dark matter.

In this paper we consider models in which the dark matter is a gauge-singlet real scalar particle $S$ that interacts with the rest of the SM through the ``Higgs portal'' via an operator of the form $SS \Phi^{\dagger} \Phi$, where $\Phi$ is the SU(2)$_L$-doublet SM Higgs field.  The singlet scalar is kept stable by imposing a symmetry $S \to -S$ on the Lagrangian.  In Sec.~\ref{sec:SM+S} we consider the minimal model in which a single gauge-singlet real scalar is added to the SM~\cite{Veltman:1989vw,Silveira:1985rk,McDonald:1993ex,Burgess:2000yq,McDonald:2001vt,Barger:2007im,Goudelis:2009zz,Gonderinger:2009jp,He:2009yd,Profumo:2010kp,Yaguna:2011qn,Drozd:2011aa,Djouadi:2011aa,Kadastik:2011aa,Djouadi:2012zc,Cheung:2012xb,Damgaard:2013kva,Cline:2013gha,Baek:2014jga,Feng:2014vea}, and show that it cannot explain the observed dark matter self-interaction cross section once we impose the requirement that the annihilation cross section of the singlet scalars into SM particles yields the correct relic abundance from thermal freeze-out together with the constraints on the invisible branching fraction of the Higgs boson from the CERN Large Hadron Collider (LHC)~\cite{ATLASinvis,CMS:2015dia}.  The correct relic abundance can, however, be achieved in this model through thermal ``freeze-in,'' in which the dark matter is produced slowly in the early universe through extremely weak interactions with the SM Higgs boson~\cite{McDonald:2001vt,Hall:2009bx,Yaguna:2011qn}.  In Sec.~\ref{sec:2HDM+S} we attempt to salvage the usual thermal freeze-out picture by considering the real singlet scalar extension of two-Higgs-doublet models (2HDMs)~\cite{He:2008qm,Grzadkowski:2009iz,Logan:2010nw,Boucenna:2011hy,He:2011gc,Bai:2012nv,He:2013suk,Cai:2013zga,Wang:2014elb,Chen:2013jvg,Drozd:2014yla,Wang:2014elb}.  We find a small viable region of parameter space in which the singlet scalar mass is tuned to be very close to half the mass of the second, as-yet-undiscovered CP-even neutral Higgs boson of these models.  In Sec.~\ref{sec:conclusions} we conclude.

%%%%%%%%%%%%%%%%%%%%%%%%%%%%%%%%%%%%%%%%%%%%%%
\section{Minimal singlet extension of the Standard Model}
\label{sec:SM+S}

We write the scalar potential for the SM plus a real singlet as
\begin{equation}
	V = -\mu^2 \Phi^{\dagger} \Phi + \frac{\mu_S^2}{2} SS +  \lambda_h (\Phi^{\dagger} \Phi)^2 
	+ \lambda_p SS \Phi^{\dagger} \Phi + \lambda_S S^4,
	\label{eq:SM+S}
\end{equation}
where we have imposed invariance under $S \to -S$ to keep the singlet stable.
The physical SM Higgs mass $m_h \simeq 125$~GeV~\cite{Aad:2015zhl} and the SM Higgs vacuum expectation value (vev) $v \equiv (\sqrt{2} G_F)^{-1/2} \simeq 246$~GeV~\cite{Agashe:2014kda} fix two of the scalar potential parameters,
\begin{eqnarray}
	\lambda_h &=& \frac{m_h^2}{2 v^2} \simeq 0.129, \nonumber \\
	\mu^2 &=& \lambda_h v^2 \simeq (88.4~{\rm GeV})^2.
\end{eqnarray}
For the potential to be bounded from below we require that $\lambda_S > 0$ and $\lambda_p > -2 \sqrt{\lambda_h \lambda_S}$.
To keep the singlet from getting a vev and thereby ceasing to be stable, we require that 
\begin{equation}
	m_S^2 = \mu_S^2 + \lambda_p v^2 > 0,
\end{equation}
where $m_S$ is the physical mass of the singlet scalar in the electroweak-breaking vacuum.
This implies that at most one of $\mu_S^2$ or $\lambda_p$ can be negative.
If $\mu_S^2 < 0$ the potential can develop a second minimum in which the singlet gets a vev and the doublet does not; to avoid this possibility we require that $\mu_S^2 > 0$.

The Feynman rules involving the singlet that will be important here are
\begin{eqnarray}
	SSSS: &&\quad -24 i \lambda_S, \nonumber \\
	SSh: &&\quad -2 i \lambda_p v.
\end{eqnarray}

\begin{figure}
\resizebox{0.2\textwidth}{!}{\includegraphics{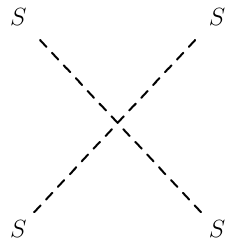}}
\resizebox{0.2\textwidth}{!}{\includegraphics{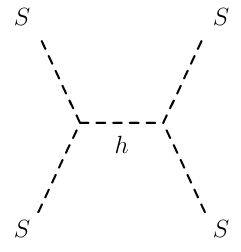}}
\resizebox{0.2\textwidth}{!}{\includegraphics{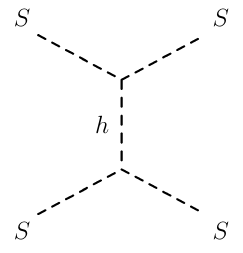}}
\resizebox{0.2\textwidth}{!}{\includegraphics{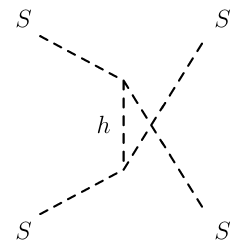}}
\caption{Feynman diagrams for the $SS \to SS$ scattering process in the real scalar singlet extension of the SM.}
\label{fig:fds}
\end{figure}

We will consider dark matter scattering $SS \to SS$ in three mass ranges.  We work in the minimal real singlet scalar extension of the SM, but what follows will inform the calculations in the extended Higgs sector that we consider later.  The relevant Feynman diagrams are shown in Fig.~\ref{fig:fds}.

\subsection{Heavy dark matter, $m_S \gg m_h$}

When $S$ is very heavy compared to the SM Higgs mass, the Higgs-exchange diagrams in Fig.~\ref{fig:fds} are suppressed by momentum factors compared to the four-point diagram, so we neglect them for simplicity.  The scattering cross section per unit dark matter mass can then be written as
\begin{equation}
	\sigma_{\rm DM}/m_S \simeq \frac{9 \lambda_S^2}{2 \pi m_S^3}.
	\label{eq:heavy}
\end{equation}

Perturbative unitarity of the $2\to 2$ scattering amplitude for $SS \to SS$ at high energy sets an upper bound on $\lambda_S$.  Requiring that the zeroth partial wave amplitude $a_0$ satisfy $|{\rm Re}~a_0| < 1/2$ and taking into account the identical particle factors of $1/\sqrt{2}$ for the initial and final states yields
\begin{equation}
	\lambda_S < 2\pi/3.
	\label{eq:lslim}
\end{equation}
Unfortunately, after imposing this constraint we find that the heavy-singlet scenario 
is inconsistent with
the measured cross section in Eq.~(\ref{eq:xsec2}).  Indeed, reproducing 
the observed dark matter self-interaction cross section would require $m_S \lesssim 0.1$~GeV; we 
will consider this low-mass scenario below.

\subsection{Dark matter near the Higgs resonance, $m_S \sim m_h/2$}
\label{sec:resonant}

When the scattering is near resonance, the $s$-channel Higgs exchange diagram in Fig.~\ref{fig:fds} dominates.  Defining the $hSS$ coupling 
\begin{equation}
	g_{SSh} = -2 \lambda_p v,
\end{equation}
we can write the $hSS$ Feynman rule as $i g_{SSh}$.  The $s$-channel Higgs exchange diagram in Fig.~\ref{fig:fds} 
then yields a dark matter self-interaction cross section per unit mass 
of\footnote{The contribution to the center-of-mass collision energy of the present-day kinetic energy 
will be totally negligible compared to the mass splitting that we will find, so we neglect it here.}
\begin{eqnarray}
	\sigma_{\rm DM}/m_S &\simeq& 
		\frac{g_{SSh}^4}{128 \pi m_S^3 ((4m_S^2 - m_h^2)^2 + \Gamma_h^2 m_h^2 ) } \nonumber \\
	&\simeq& \frac{v^4 \lambda_p^4}{16 \pi m_h^5 ((m_S - m_h/2)^2 +\Gamma_h^2/4)},
	\label{eq:midrange}
\end{eqnarray}
where $\Gamma_h$ is the Higgs boson width and in the last expression we have taken $m_S \simeq m_h/2$. 

Perturbative unitarity of the four $SS \to \phi_i \phi_i$ scattering amplitudes, where $\phi_i$ 
are the four real scalar components of the SM Higgs doublet, $\Phi$, constrains
\begin{equation}
| \lambda_p | < 4\pi,  \qquad \hbox{or} \qquad |g_{SSh} | < 8\pi v. 
\end{equation}
The region $m_S < m_h/2$,  because of the contribution of the $h\to SS$ channel to the Higgs width,
is severely constrained such that $m_h/2 - m_S \lesssim 10^{-6}$~GeV in order to achieve the observed
DM  self-interaction cross section and for the invisible Higgs width to be consistent with experiment.
For the region above half the Higgs mass, this decay channel is not kinematically allowed, and the restriction
reduces to $m_S - m_h/2 \lesssim 0.11$~GeV, where we use the lower value of $\sigma_{DM}/m_S$
in Eq.~(\ref{eq:xsec2} to obtain the upper bound.  This represents a tuning of about two parts per mille.

We now show that this scenario cannot be reconciled with the observed dark matter relic density because the annihilation cross section into SM particles through the Higgs resonance is several orders of magnitude too large.

\subsubsection{Relic density}

The annihilation cross section required to obtain the correct thermal dark matter relic density for $m_S \sim m_h/2$ is~\cite{Steigman:2012nb}
\begin{equation}
	\sigma v_{\rm rel} \simeq 2.2 \times 10^{-26}~{\rm cm}^3~{\rm s}^{-1}
	\simeq 1.9 \times 10^{-9}~{\rm GeV}^{-2}.
	\label{eq:thermalxsec}
\end{equation}

\begin{figure}
\resizebox{0.2\textwidth}{!}{\includegraphics{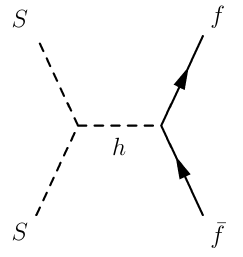}}
\resizebox{0.2\textwidth}{!}{\includegraphics{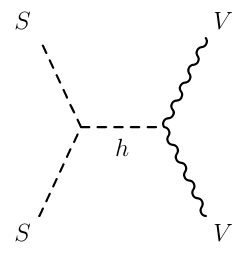}}
\caption{Feynman diagrams for the annihilation of singlet scalars into SM particles in the real singlet scalar extension of the SM.}
\label{fig:anni}
\end{figure}

For $m_S \sim m_h/2$, the singlet scalars annihilate into SM particles predominantly through $s$-channel Higgs exchange, as shown in Fig.~\ref{fig:anni}.  The temperature of the thermal bath at the time of freeze-out for a dark matter particle with the correct relic density is $T \sim m_S/20 \sim 5$~GeV~\cite{Gelmini:2015zpa}.  This is much larger than the mass splitting $|m_S - m_h/2| \lesssim 0.11$~GeV required to obtain a large enough dark matter self-interaction cross section, so thermal effects during freeze-out cannot be neglected.

The annihilation cross section can be written as 
\begin{equation}
	\sigma v_{\rm rel} = \frac{2 g_{SSh}^2 \Gamma^{h_{\rm SM}}_{\rm tot}}{E_{\rm cm} (E_{\rm cm}^2 - m_h^2)^2} ,
	\label{eq:sigvrel}
\end{equation}
where $E_{\rm cm} = 2 m_S + m_S v_{\rm rel}^2/4$ is the collision center-of-mass energy, $v_{\rm rel}$ is the relative velocity of the two dark matter particles in the center-of-momentum frame and $\Gamma^{h_{\rm SM}}_{\rm tot} \simeq 4.1$~MeV is the SM Higgs total width~\cite{Heinemeyer:2013tqa}.  

The singlet scalar can be heavier or lighter than half the Higgs mass.  For a singlet scalar heavier than half the Higgs mass, the thermal motion pushes the annihilation process farther from resonance.  Imposing the self-interaction cross section measurement and integrating the annihilation cross section over the Boltzmann distribution with $T \sim m_S/20$ yields
\begin{equation}
	\langle \sigma v_{\rm rel} \rangle \gtrsim 0.03~{\rm GeV}^{-2},
\end{equation}
where the bound is saturated for $|g_{SSh}|$ as large as possible.
This is many orders of magnitude larger than the required cross section for thermal freeze-out given in Eq.~(\ref{eq:thermalxsec}), resulting in a singlet scalar relic abundance way too small to account for the observed dark matter.  If the singlet scalar is lighter than half the Higgs mass, the thermal motion pushes the annihilation process onto the resonance resulting in an even larger thermally-averaged cross section and even smaller relic abundance.  These large cross sections are a consequence of the large $SSh$ coupling and the tuning of the singlet mass close to the Higgs resonance that is required to achieve the observed dark matter self-interaction cross section.

%%%%%%%%%%%%%%%%%%%%%%%%%%%%%%%%%%%%%%%%%%%%%%
\subsection{Low-mass dark matter, $m_S \ll m_h$}
\label{sec:lightdm}

\begin{figure}
\resizebox{0.5\textwidth}{!}{\includegraphics{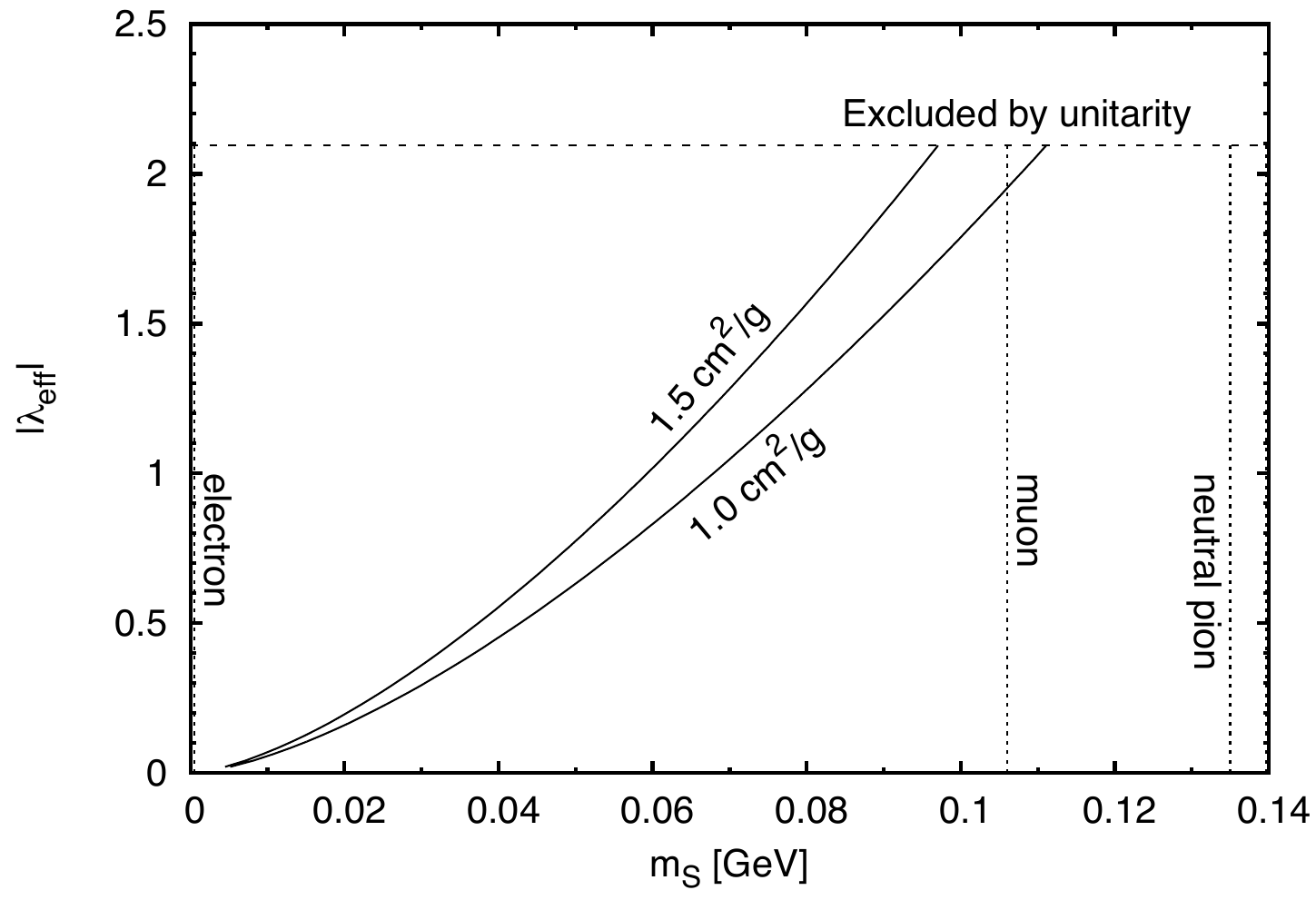}}
\caption{Favored region in $m_S$ and the effective $4S$ coupling $\lambda_{\rm eff}$ defined in Eq.~(\ref{eq:lambdaeff}) for a dark matter self-interaction cross section in the range 1.0--1.5~cm$^2$~g$^{-1}$.  Values of $|\lambda_{\rm eff}|$ above $2 \pi/3 \simeq 2.094$ (horizontal dashed line) are excluded by perturbative unitarity considerations.  The vertical dotted lines show the thresholds for $SS$ annihilation into pairs of electrons, muons, and neutral pions.  The charged pion threshold is at the right edge of the plot.}
\label{fig:plot}
\end{figure}

We now consider the situation of low-mass dark matter.  In this case all four diagrams in Fig.~\ref{fig:fds} contribute.  At low center-of-mass energies, the Higgs boson $h$ can be integrated out, yielding an effective four-$S$ coupling with Feynman rule $-24 i \lambda_{\rm eff}$, where
\begin{equation}
	\lambda_{\rm eff} = \lambda_S - \frac{g_{SSh}^2}{8 m_h^2}.
	\label{eq:lambdaeff}
\end{equation}
$SS \to SS$ scattering in the low-energy effective theory must also satisfy perturbative unitarity, so as in Eq.~(\ref{eq:lslim}) we obtain
\begin{equation}
	|\lambda_{\rm eff}| < 2 \pi/3.
\end{equation}
The dark matter scattering cross section per unit mass as in Eq.~(\ref{eq:heavy}) becomes
\begin{equation}
	\sigma_{\rm DM}/m_S \simeq \frac{9 \lambda_{\rm eff}^2}{2 \pi m_S^3}.
\end{equation}
The range of $m_S$ and $\lambda_{\rm eff}$ consistent with the cross section measurement from Ref.~\cite{Kahlhoefer:2015vua} [Eq.~(\ref{eq:xsec2})] is shown in Fig.~\ref{fig:plot}.  In particular, the singlet scalar dark matter must be very light, with mass below about 0.1~GeV.

%This scenario is unconstrained by dark matter direct-detection limits because of the very low dark matter mass, which leads to very low energy transfers to the target nuclei in scattering events.  The most sensitive limits for low-mass dark matter currently come from a special run of the SuperCDMS experiment (known as ``CDMS-lite''), which probed dark matter particle masses down to about 3~GeV~\cite{Agnese:2013jaa} (see also Ref.~\cite{Agnese:2014aze}).

\subsubsection{Invisible Higgs decays}

When $m_S \ll m_h$, the non-observation of invisible Higgs boson decays constrains $\lambda_p$.

The partial width for $h \to SS$ is
\begin{equation}
	\Gamma(h \to SS) = \frac{g_{SSh}^2}{32 \pi m_h} \sqrt{1 - \frac{4 m_S^2}{m_h^2}}
	\simeq \frac{g_{SSh}^2}{32 \pi m_h},
\end{equation}
where the last expression holds for $m_S \ll m_h$.
The invisible Higgs branching fraction is defined as
\begin{equation}
	{\rm BR}(h \to {\rm invis.}) = \frac{\Gamma(h \to SS)}
		{\Gamma^{h_{\rm SM}}_{\rm tot} + \Gamma(h \to SS)}.
\end{equation}

The strongest constraint on ${\rm BR}(h \to {\rm invis.})$ currently comes from the ATLAS search for invisible Higgs decays in vector boson fusion production, with a 95\% confidence level upper limit of ${\rm BR}(h \to {\rm invis.}) \leq 0.29$ for a 125~GeV Higgs boson, assuming a SM production cross section~\cite{ATLASinvis}.  This leads to an upper bound on $g_{SSh}^2$, or equivalently $|\lambda_p|$, of
\begin{equation}
	g_{SSh}^2 \leq 21~{\rm GeV}^2, \qquad {\rm or} \quad
	|\lambda_p| \leq 9.3 \times 10^{-3}.
\end{equation}
This constraint is very strong compared to the perturbative unitarity constraint $|\lambda_p| < 4 \pi$.

One can also derive an upper bound on $g_{SSh}^2$ from the measured signal strengths $\mu_i$ of the Higgs boson in various SM decay channels.  In the singlet scalar extension of the SM, these signal strengths are all modified by the common factor
\begin{equation}
	\mu_i \equiv \mu = \frac{\Gamma_{\rm tot}^{h_{\rm SM}}}
	{\Gamma_{\rm tot}^{h_{\rm SM}} + \Gamma(h \to SS)}.
\end{equation}
Combining all channels, ATLAS~\cite{ATLASmu} and CMS~\cite{Khachatryan:2014jba} respectively measure an overall signal strength of
\begin{eqnarray}
	\mu &=& 1.18^{+0.15}_{-0.14}~~{\rm (ATLAS)}, \nonumber \\
	\mu &=& 1.00^{+0.14}_{-0.13}~~{\rm (CMS)}.
\end{eqnarray}
In the real singlet extension of the SM, $\mu \leq 1$, therefore, relative to this best-fit point, the $\Delta \chi^2 = 4$ lower bounds on the signal strength are
\begin{eqnarray}
	\mu &\geq& 0.85~~{\rm (ATLAS)}, \nonumber \\
	\mu &\geq& 0.74~~{\rm (CMS)}.
\end{eqnarray}
These in turn lead to upper bounds on $g_{SSh}^2$, or equivalently $|\lambda_p|$, of
\begin{eqnarray}
	&&g_{SSh}^2 \leq 9.1~{\rm GeV}^2, \quad {\rm or} \quad |\lambda_p| \leq 6.1 \times 10^{-3}~~{\rm (ATLAS)}, \nonumber \\
	&&g_{SSh}^2 \leq 18~{\rm GeV}^2, \, \quad {\rm or} \quad |\lambda_p| \leq 8.6 \times 10^{-3}~~{\rm (CMS)}.
\end{eqnarray}

\subsubsection{Relic density: thermal freeze-out}

The annihilation cross section required to obtain the correct thermal dark matter relic density varies slightly with $m_S$~\cite{Steigman:2012nb}.  For $m_S$ around 0.1~GeV, the annihilation cross section needs to be~\cite{Steigman:2012nb}
\begin{equation}
	\sigma v_{\rm rel} \simeq 4.8 \times 10^{-26}~{\rm cm}^3~{\rm s}^{-1}
	\simeq 4.1 \times 10^{-9}~{\rm GeV}^{-2}.
	\label{eq:sigvdesired}
\end{equation}

In the low-mass range, the singlet scalars annihilate into SM particles predominantly through  $s$-channel Higgs exchange, as shown in Fig.~\ref{fig:anni}.  This annihilation cross section can be conveniently expressed for $m_S \ll m_h$ as
\begin{equation}
	\sigma v_{\rm rel} \simeq \frac{4 v^2 \lambda_p^2}{m_h^4} 
	\frac{\Gamma^{h_{\rm SM}}_{\rm tot}(m = 2 m_S)}{m_S},
	\label{eq:smsigv}
\end{equation}
where $\Gamma^{h_{\rm SM}}_{\rm tot}(m = 2 m_S)$ is the would-be total width of the SM Higgs boson if its mass were equal to $2 m_S$.  The largest possible values of $\sigma v_{\rm rel}$ are obtained at the high end of the singlet mass range shown in Fig.~\ref{fig:plot}, when annihilations into muon pairs are kinematically accessible.  In that case, $\Gamma^{h_{\rm SM}}_{\rm tot}(m = 2m_S)$ is given to a good approximation by the decay width into muon pairs,
\begin{equation}
	\Gamma(h^* \to \mu\mu) = \frac{1}{8 \pi} \frac{m_{\mu}^2}{v^2} (2 m_S) 
	\left[ 1 - \frac{m_{\mu}^2}{m_S^2} \right]^{3/2}.
\end{equation}
Inserting this into Eq.~(\ref{eq:smsigv}) yields
\begin{equation}
	\sigma v_{\rm rel} \simeq (1.5 \times 10^{-11}~{\rm GeV}^{-2}) \, \lambda_p^2 
	\left[ 1 - \frac{m_{\mu}^2}{m_S^2} \right]^{3/2}.
\end{equation}
Even ignoring the kinematic suppression from the muon mass and the upper bound on $|\lambda_p|$ from LHC Higgs data, it is not possible to obtain a large enough dark matter annihilation cross section while satisfying the constraint on $\lambda_p$ from perturbative unitarity; one would need $|\lambda_p| \simeq 5 \pi$, in violation of Eq.~(\ref{eq:lpuni}).  Imposing the constraint on $|\lambda_p|$ from LHC searches for invisible Higgs decays or the measurement of the Higgs signal strengths yields an annihilation cross section more than six orders of magnitude too small to yield the correct thermal freeze-out dark matter relic abundance.

We also note that, even if the appropriate annihilation cross section for thermal freeze-out could be obtained, this low-mass scenario with $SS \to \mu\mu$ is strongly excluded by cosmic microwave background constraints arising from the electromagnetic energy injection at the time of recombination~\cite{Lopez-Honorez:2013cua} (for an update with Planck data see Ref.~\cite{Ade:2015xua}).

\subsubsection{Relic density: thermal freeze-in}

The singlet scalar dark matter extension of the SM can easily account for the observed dark matter self interactions if the singlet mass is below about 0.1~GeV, as shown in Fig.~\ref{fig:plot}.  The model fails when it attempts to account for the dark matter relic abundance via thermal freeze-out, because the singlet annihilation cross section into SM particles cannot be made large enough.  However, there is an alternative perfectly viable way to obtain the correct dark matter relic abundance through thermal \emph{freeze-in}~\cite{McDonald:2001vt,Hall:2009bx}.  In this scenario, the cross section for two dark matter particles to annihilate into two SM particles, or vice versa, is extremely small, so that the dark matter never comes into thermal equilibrium with the SM plasma in the early universe.  Instead, dark matter is slowly produced through annihilations of SM particles, until the universe becomes too cool for dark matter to be produced.  The resulting dark matter abundance ``freezes in'' and persists to the present day.  Because annihilation of dark matter particles never becomes important, the kinematic thresholds shown in Fig.~\ref{fig:plot} are irrelevant in this scenario.

This scenario was studied for the singlet scalar extension of the SM in Refs.~\cite{McDonald:2001vt,Yaguna:2011qn}.  In our scenario with $m_S \lesssim 0.1$~GeV, the correct dark matter relic abundance is obtained for $|\lambda_p| \sim (2-3) \times 10^{-11}$~\cite{Yaguna:2011qn}.\footnote{Note that, for this coupling strength, the $\lambda_p SS \Phi^{\dagger} \Phi$ term contributes about $(1~{\rm MeV})^2$ to $m_S^2$~\cite{McDonald:2001vt,Kang:2014cia}.}  This is easily consistent with the upper bounds $|\lambda_p| < (6.1 - 9.3) \times 10^{-3}$ from LHC Higgs signal strength measurements and invisible decay branching ratio limits.  Such a small $SS \Phi^{\dagger} \Phi$ interaction strength is even technically natural: if set to zero, $\lambda_p$ cannot be regenerated through loops involving other couplings in the model.  The only disadvantage is phenomenological: the dark matter becomes impossible to detect through non-gravitational interactions with SM particles.

%%%%%%%%%%%%%%%%%%%%%%%%%%%%%%%%%%%%%%%%%%%%%%
\section{Singlet extension of the two Higgs doublet model}
\label{sec:2HDM+S}

To try to salvage the thermal freeze-out scenario for singlet scalar dark matter, we consider extended models with two resonances, the discovered SM-like Higgs boson $h$ with mass 125~GeV and a second CP-even scalar $H$, which can be lighter or heavier than $h$.  Generic models allow both states to couple to $SS$; as a prototype we consider the singlet scalar dark matter extension of the two-Higgs-doublet model (2HDM)~\cite{He:2008qm,Grzadkowski:2009iz,Logan:2010nw,Boucenna:2011hy,He:2011gc,Bai:2012nv,He:2013suk,Cai:2013zga,Wang:2014elb,Chen:2013jvg,Drozd:2014yla,Wang:2014elb}.

As mentioned previously, thermal freeze-out in the low-mass scenario with $m_S \lesssim 0.1$~GeV is already strongly excluded by cosmic microwave background constraints~\cite{Lopez-Honorez:2013cua}.  We therefore focus in what follows on resonant scattering and annihilation through the second scalar resonance $H$.  
In the singlet scalar extension of the SM, the resonant-scattering scenario failed because the cross section for dark matter annihilation via the Higgs resonance was several orders of magnitude too large.  In the singlet scalar extension of the 2HDM, we can get around this by having the dark matter interact resonantly through the $H$ pole while making the coupling of $H$ to SM particles sufficiently weak, thereby suppressing the thermal dark matter annihilation cross section through the $H$ resonance.  
As we will see, this will not work in the usual 2HDMs with natural flavor conservation, but it can be made to work in the Yukawa-aligned 2HDM extended with a singlet.

Of course, the singlet scalar extension of the 2HDM can also account for the observed dark matter relic density through thermal \emph{freeze-in} when $m_S \lesssim 0.1$~GeV and the couplings of $S$ to the two Higgs doublets are both extremely small.

%%%%%%%%%%%%%%%%%%%%%%%%%%%%%%%%%%%%%%%%%%%%%%
\subsection{2HDM with natural flavor conservation}

The scalar potential for the singlet scalar dark matter extension of the 2HDM can be written as
\begin{equation}
	V = \frac{\mu_S^2}{2} S^2 + \lambda_S S^4 + \lambda_{p1} SS \Phi_1^{\dagger} \Phi_1
	+ \lambda_{p2} SS \Phi_2^{\dagger} \Phi_2 + V_{\rm 2HDM},
	\label{eq:V}
\end{equation}
where $\Phi_1$ and $\Phi_2$ are the two Higgs doublets and $V_{\rm 2HDM}$ is the usual scalar potential of the 2HDM.  We have imposed a second $Z_2$ symmetry under which $\Phi_1 \to -\Phi_1$, which can be softly broken by a dimension-two term in $V_{\rm 2HDM}$.
This allows us to avoid flavor-changing neutral Higgs interactions through natural flavor conservation~\cite{Glashow:1976nt}, by forcing the right-handed fermions to couple to only one Higgs doublet.  The transformation properties of the right-handed fermions under the second $Z_2$ symmetry determine the ``Type'' of 2HDM (for a review see Ref.~\cite{Branco:2011iw}).

To set the notation, we write the vacuum expectation values as 
\begin{equation}
	\langle \Phi_1 \rangle = \left( \begin{array}{c} 0 \\ v_1/\sqrt{2} \end{array} \right), \qquad \qquad
	\langle \Phi_2 \rangle = \left( \begin{array}{c} 0 \\ v_2/\sqrt{2} \end{array} \right),
\end{equation}
with $v_1^2 + v_2^2 = v^2 \simeq (246~{\rm GeV})^2$ and $\tan\beta \equiv v_2/v_1$.  In terms of the real neutral components of the two doublets $\Phi_i^0 \equiv \phi^{0,r}_i/\sqrt{2}$, the physical CP-even neutral scalars $h$ (identified as the 125~GeV Higgs) and $H$ are defined as
\begin{eqnarray}
	h &=& -\sin\alpha \, \phi_1^{0,r} + \cos\alpha \, \phi_2^{0,r}, \nonumber \\
	H &=& \cos\alpha \, \phi_1^{0,r} + \sin\alpha \, \phi_2^{0,r},
\end{eqnarray}
where $\alpha$ is a mixing angle.

In terms of these parameters, the Feynman rules for the $SSh$ and $SSH$ vertices are $i g_{SSh}$ and $i g_{SSH}$, respectively, with
\begin{eqnarray}
	g_{SSh} &=& 2 v ( \lambda_{p1} \cos\beta \sin\alpha - \lambda_{p2} \sin\beta \cos\alpha), 
	\nonumber \\
	g_{SSH} &=& 2 v ( -\lambda_{p1} \cos\beta \cos\alpha - \lambda_{p2} \sin\beta \sin\alpha).
	\label{eq:2hdmcoups}
\end{eqnarray}
Perturbative unitarity of the $SS \to \Phi_1\Phi_1$ and $SS \to \Phi_2\Phi_2$ scattering amplitudes separately constrain\footnote{A proper coupled-channel analysis would lead to a tighter constraint.}
\begin{equation}
	|\lambda_{p1} | < 4 \pi, \qquad \qquad |\lambda_{p2}| < 4 \pi,
	\label{eq:2hdmuni}
\end{equation}
so $|g_{SSH}|$ can be as large as $8 \pi v$.

In the 2HDM plus a singlet scalar, the low-energy effective $SS \to SS$ coupling for $m_S \ll m_h$, $m_H$ is given by
\begin{equation}
	\lambda_{\rm eff} = \lambda_S - \frac{g_{SSh}^2}{8 m_h^2} - \frac{g_{SSH}^2}{8 m_H^2}.
\end{equation}
The constraint $m_S \lesssim 0.1$~GeV carries through from the analysis in the previous section, so we do not pursue this low-mass scenario further.

To avoid experimental constraints from $h$ coupling measurements, we tune $\sin(\beta - \alpha) \simeq 1$.  This means that
\begin{equation}
	\sin\alpha \simeq -\cos \beta, \qquad \qquad
	\cos\alpha \simeq \sin\beta.
	\label{eq:sba}
\end{equation}
For $\sin(\beta-\alpha) \simeq 1$, the $H$ coupling to vector boson pairs is heavily suppressed, being proportional to $\cos(\beta-\alpha)$.  This allows $H$ to evade searches for invisibly-decaying Higgs bosons via $H$ production in vector boson fusion or associated $VH$ production.  We focus on the Type-I 2HDM, since in that model all the fermion Yukawa couplings can be suppressed by making $\cot\beta$ small.  The $H$ couplings to fermions in the Type-I 2HDM are equal to the corresponding couplings of the SM Higgs boson multiplied by a common scaling factor
\begin{equation}
	\kappa_f^H = \frac{\sin\alpha}{\sin\beta} \simeq - \cot\beta,
	\label{eq:kappaf}
\end{equation}
where we used Eq.~(\ref{eq:sba}) to obtain the coupling in the $\sin(\beta - \alpha) \to 1$ limit.
Perturbativity of the top quark Yukawa coupling requires $\cot\beta < 10/3$~\cite{Barger:1989fj}.

For $m_S \sim m_H/2$, the $SS \to SS$ scattering is dominated by the $s$-channel $H$ exchange diagram, yielding a cross section per unit dark matter mass
\begin{equation}
	\sigma_{\rm DM}/m_S \simeq \frac{g_{SSH}^4}{128 \pi m_S^3 (4m_S^2 - m_H^2)^2}.
\end{equation}
As in the case of the singlet scalar extension of the SM, the singlet mass must be tuned to be extremely close to $m_H/2$ for this scattering cross section to account for the halo displacement observed in Ref.~\cite{Massey:2015oza}.

The dark matter annihilation cross section to SM particles, which controls the relic density, is similarly given for annihilation through the $H$ resonance by
\begin{equation}
	\sigma v_{\rm rel} \simeq \frac{2 g_{SSH}^2 \Gamma_{\rm tot}^{H \to {\rm SM}}}
		{E_{\rm cm} (E_{\rm cm}^2 - m_H^2)^2}.
	\label{eq:sigvyukal}
\end{equation}
These expressions take an analogous form to those in Eqs.~(\ref{eq:midrange}) and (\ref{eq:sigvrel}) for the singlet scalar extension of the SM.  In that model, the problem was that the annihilation cross section was much too large once the dark matter self-interaction cross section was fixed to the newly-observed value.  We therefore attempt to suppress the width of $H$ into SM particles while keeping $g_{SSH}$ large.\footnote{Note that, if $m_H < m_h$, the $h \to SS$ invisible decay becomes kinematically accessible and must be avoided by tuning $g_{SSh}$ to be small.  Together with Eq.~(\ref{eq:sba}), this implies that
\begin{equation}
	\lambda_{p1} \cot \beta \simeq -\lambda_{p2} \tan \beta,
	\label{eq:l1l2}
\end{equation}
and hence 
\begin{equation}
	g_{SSH} \simeq 2 v \lambda_{p1} \cot\beta \simeq -2 v \lambda_{p2} \tan\beta.
	\label{eq:gSSH}
\end{equation}}

We take $m_H > m_h$, so that $m_S > m_h/2$ and there is no need to tune $g_{SSh}$ to be small, since invisible decays $h \to SS$ will be kinematically forbidden.
To suppress $H \to WW, ZZ$ and keep the $h$ couplings SM-like, we require $\sin(\beta-\alpha) \to 1$.  To suppress $H$ decays to fermions, we take $\cot\beta \ll 1$ in the Type-I 2HDM.  Then the decay width of $H$ to SM particles, for $m_H < 2 m_t$, can be approximated as
\begin{equation}
	\Gamma_{\rm tot}^{H \to {\rm SM}} \simeq (\kappa_f^H)^2 \left(\frac{m_H}{125~{\rm GeV}}\right)
	[1 - {\rm BR}(h_{\rm SM} \to VV)] \, \Gamma_{\rm tot}^{h_{\rm SM}},
	\label{eq:gam2HDM}
\end{equation}
where ${\rm BR}(h_{\rm SM} \to VV) \equiv {\rm BR}(h_{\rm SM} \to WW) + {\rm BR}(h_{\rm SM} \to ZZ) \simeq 0.24$~\cite{Heinemeyer:2013tqa}.  In particular, $\Gamma_{\rm tot}^{H \to {\rm SM}} \propto (\kappa_f^H)^2 \simeq \cot^2\beta$.

Comparing Eq.~(\ref{eq:sigvrel}) with Eqs.~(\ref{eq:sigvyukal}) and (\ref{eq:gam2HDM}), we see that we need to severely suppress the parameter combination $(\kappa_f^H)^2/g_{SSH}^2$.
However, in this model the $SSH$ coupling is
\begin{equation}
	g_{SSH} \simeq 2 v \sin\beta \cos\beta (-\lambda_{p1} + \lambda_{p2}),
\end{equation}
where we have used $\sin(\beta - \alpha) \simeq 1$.
In particular, $g_{SSH} \propto \cos\beta$, so that the ratio $(\kappa_f^H)^2 / g_{SSH}^2$ is not at all suppressed at $\cot\beta \ll 1$.  Indeed, once the upper bounds $|\lambda_{p1}| < 4\pi$ and $|\lambda_{p2}| < 4\pi$ from perturbative unitarity are imposed, the situation is no better than in the singlet scalar extension of the SM discussed in Sec.~\ref{sec:resonant}.

%%%%%%%%%%%%%%%%%%%%%%%%%%%%%%%%%%%%%%%%%
\subsection{Yukawa-aligned 2HDM}

In the preceding section our main problem was that suppressing the $H$ couplings to fermions simultaneously suppressed the $SSH$ coupling, so that the factor $(\kappa_f^H)^2/g_{SSH}^2$ in Eq.~(\ref{eq:sigvyukal}) remained no smaller than in the singlet scalar extension of the SM.  This could be avoided if the scalar potential contained a term $(SS \Phi_1^{\dagger} \Phi_2 + {\rm h.c.})$.  Such a term breaks the $Z_2$ symmetry imposed to avoid flavor-changing neutral Higgs interactions; however, there is no reason to assume a priori that flavor conservation originates from the structure of the scalar potential.

The Yukawa-aligned 2HDM~\cite{Pich:2009sp} allows both Higgs doublets to couple to all fermions, and avoids flavor-changing neutral Higgs interactions by positing that the (unknown) flavor physics causes the two Yukawa matrices for the up-type quarks to be proportional to each other, and similarly for the down-type quarks and the charged leptons.  Then one can add a third term to the scalar potential in Eq.~(\ref{eq:V}) coupling the singlet scalar to the doublets,
\begin{equation}
	V \supset (\lambda_{p3} S S \Phi_1^{\dagger} \Phi_2 + {\rm h.c.}),
	\label{eq:yukal}
\end{equation}
where for simplicity we will avoid CP violation by choosing $\lambda_{p3}$ to be real.\footnote{The singlet scalar extension of the Yukawa-aligned 2HDM was considered in Ref.~\cite{Wang:2014elb}, but the term in Eq.~(\ref{eq:yukal}) was not considered there.}  The couplings in Eq.~(\ref{eq:2hdmcoups}) become
\begin{eqnarray}
	g_{SSh} &=& 2 v [ \lambda_{p1} \cos\beta \sin\alpha - \lambda_{p2} \sin\beta \cos\alpha
	\nonumber \\
	&& - \lambda_{p3} (\cos\beta \cos\alpha - \sin\beta \sin\alpha) ], 
	\nonumber \\
	g_{SSH} &=& 2 v [ -\lambda_{p1} \cos\beta \cos\alpha - \lambda_{p2} \sin\beta \sin\alpha 
	\nonumber \\
	&& - \lambda_{p3} (\cos\beta \sin\alpha + \sin\beta \cos\alpha)].
\end{eqnarray}

Tuning $\sin(\beta - \alpha) \simeq 1$ to make the $h$ couplings SM-like reduces these to
\begin{eqnarray}
	g_{SSh} &\simeq& 2 v [ -\lambda_{p1} \cos^2\beta - \lambda_{p2} \sin^2\beta - 2 \lambda_{p3} \sin\beta \cos\beta ],
	\nonumber \\
	g_{SSH} &\simeq& 2 v [ (-\lambda_{p1} + \lambda_{p2}) \sin\beta \cos\beta  \nonumber \\
	&& - \lambda_{p3} (\sin^2\beta - \cos^2\beta)].
\end{eqnarray}

Now a suppression of the $(\kappa_f^H)^2$ factor in Eq.~(\ref{eq:sigvyukal}) need not be counteracted by a suppression of $g_{SSH}^2$.
We can choose the relative Yukawa couplings of the two Higgs doublets so that the $H$ couplings to fermions are all suppressed while at the same time keeping $g_{SSH}$ unsuppressed.  We need a fermionic coupling suppression $\kappa_f^H \sim 10^{-4}$ in order to obtain an annihilation cross section in the right range for thermal freeze-out.  For a full treatment one should also take into account the effect of thermal averaging of the annihilation cross section at the time of freeze-out, which may be important due to the severe tuning of the singlet mass relative to the $H$ resonance.  We leave this for future work.  

We finally remark that, if $m_H > m_h$, this scenario offers the possibility of direct dark matter detection due to scattering through exchange of $h$.  The $SSh$ coupling need not be small if $h \to SS$ is kinematically forbidden.  However, the couplings of $H$ to fermions (and hence to the nucleons) must be severely suppressed.  Thus the direct detection cross section in this scenario is controlled by different model parameters than the annihilation cross section, so the two are not necessarily correlated.  Therefore a direct detection signal in this model would not necessarily probe the coupling responsible for the relic density.

%%%%%%%%%%%%%%%%%%%%%%%%%%%%%%%%%%%%%%%%%%%%%%
\section{Conclusions}
\label{sec:conclusions}

In this paper we use the apparent observation of dark matter self-interactions in the cluster Abell 3827 to constrain models of singlet scalar dark matter.  We showed that the most natural scenario in this class of models is very light dark matter, below about 0.1~GeV, whose relic abundance is set by \emph{freeze-in} through extraordinarily weak interactions with the Higgs doublet(s) through a quartic coupling of order $10^{-11}$.

Thermal freeze-out can be salvaged by extending the Higgs sector with another doublet, so long as flavor conservation is achieved through Yukawa alignment.  The solution is, however, quite fine tuned: the singlet mass must be tuned very close (of order a part per mille) to half the mass of the second CP-even neutral Higgs boson.

%%%%%%%%%%%%%%%%%%%%%%%%%%%%%%%%%%%%%%%%%%%%%%
\begin{acknowledgments}
This work was supported by the Natural Sciences and Engineering Research Council of Canada. 
The authors thank Jim Cline for helpful conversations. \end{acknowledgments}
%%%%%%%%%%%%%%%%%%%%%%%%%%%%%%%%%%%%%%%%%%%%%%

%%%%%%%%%%%%%%%%%%%%%%%%%%%%%%%%%%%%%%%%%%%%%

\end{document}